# Long-Lived Interlayer Excitons and Type-II Band Alignment in Janus MoTe$_2$/CrSBr van der Waals Heterostructures

Mohammad Ali Mohebpour[1], Peter C Sherrell[2], Catherine Stampfl[3], Carmine Autieri[4], and Meysam Bagheri Tagani[1,4,] *

[1]Department of Physics, University of Guilan, P. O. Box 41335-1914, Rasht, Iran
[2]School of Science, RMIT University, Melbourne, VIC 3001, Australia
[3]School of Physics, The University of Sydney, Sydney NSW 2006, Australia
[4]International Research Centre Magtop, Institute of Physics, Polish Academy of Sciences, Aleja Lotników 32/46, 02668 Warsaw, Poland

## Abstract

Identifying two-dimensional heterostructures with exceptional electronic and optical properties remains an active area of research in advanced optoelectronics. Here, we present a comprehensive first-principles investigation of the electronic, optical, and excitonic properties of a MoTe$_2$/CrSBr van der Waals heterostructure using density functional theory combined with fully relativistic GW and Bethe-Salpeter equation calculations. The close lattice matching between the two monolayers enables the formation of stable heterobilayers with two inequivalent interfaces (Te-S and Te-Br) arising from the Janus nature of CrSBr. Both interfaces are dynamically and thermally stable and exhibit type-II band alignment with a direct quasiparticle gap, promoting efficient spatial separation of electrons and holes. The heterostructure hosts interlayer excitons with lifetimes (18 – 45 ps) significantly longer than those of the intralayer excitons in the isolated MoTe$_2$ (3.6 ps) and CrSBr (8.1 ps) monolayers. Moreover, the optical gap, exciton binding energy, and exciton lifetime of the heterostructure are strongly modulated by the built-in electric field associated with the Janus layer. These results establish the MoTe$_2$/CrSBr heterostructure as a versatile platform for engineering long-lived interlayer excitons and highlight its potential for next-generation optoelectronic and light-harvesting applications.

# I. INTRODUCTION

Van der Waals (vdW) heterostructures have attracted significant attention in recent years due to the emergence of electronic and optical phenomena that do not exist in isolated monolayers [1-3]. Despite the emergence of new characteristics, the weak interlayer coupling largely preserves the intrinsic properties of the individual layers [4]. By vertically stacking different two-dimensional (2D) materials, key properties of these heterostructures such as the band gap [5], band alignment [6-8], carrier mobility [9], carrier distribution [10, 11], and excitonic response [12, 13] can be systematically tailored. In particular, the relative band alignment between the layers can be engineered through material selection and stacking configuration [14, 15]. Heterostructures exhibiting type-II band alignment are especially important, as they promote the spatial separation of electrons and holes across different layers, which suppresses recombination and facilitates the formation of long-lived interlayer excitons [16]. Such control over exciton dynamics is crucial for the development of efficient optoelectronic and light-harvesting devices [17].

Among 2D semiconductors, transition metal dichalcogenides (TMDs) have been extensively studied owing to their tunable band gaps [18, 19], strong spin-orbit coupling (SOC) [20, 21], and optical response dominated by tightly bound excitons [22-24]. These characteristics make TMDs promising candidates for next-generation optoelectronic applications [25-27]. However, in isolated TMD monolayers, excitons are confined within a single layer and offer limited tunability. These intralayer excitons exhibit relatively short lifetimes due to the strong spatial overlap between electrons and holes [28], which enhances radiative recombination. In contrast, when TMDs are incorporated into vdW heterostructures with type-II band alignment, electrons and holes can reside in different layers [29-33]. This spatial separation reduces the recombination rate and gives rise to long-lived interlayer excitons [34, 35], which are highly desirable for applications requiring efficient charge separation and extended carrier lifetimes, such as photodetectors and solar energy conversion [36].

An additional and powerful strategy for property engineering in vdW heterostructures is the incorporation of Janus monolayers [37, 38]. Owing to their broken out-of-plane symmetry, Janus materials possess intrinsic dipoles and built-in electric fields that can significantly modify interfacial electrostatic potentials [39]. When integrated into a vdW heterostructure, this polarity naturally creates different interfaces depending on the stacking orientation, offering an intrinsic mechanism to tune the band gap, optical gap, exciton binding energy, and exciton lifetime without the need for external fields [40-42]. Despite all advances, the role of Janus-induced intrinsic electric fields in modulating exciton binding energies and radiative lifetimes in TMD-based heterostructures remains insufficiently understood.

In this context, combining a non-Janus TMD monolayer with a Janus monolayer provides a particularly compelling platform. The MoTe$_2$ monolayer is a stable and experimentally realized TMD [43] that exhibits strong SOC [44, 45] and noticeable excitonic effects [45, 46], yet it is not intrinsically spin-polarized [47]. In contrast, the CrSBr monolayer is spin-polarized [48] and possesses an intrinsic out-of-plane electric field arising from its Janus structure [49], while showing a comparatively weak response to SOC. Importantly, the two monolayers have closely matched lattice constants, enabling the formation of a high-quality heterostructure with minimal strain [50]. These complementary characteristics create a heterobilayer in which spin polarization and an intrinsic electric field coexist, offering enhanced opportunities to control quasiparticle and excitonic behavior. Furthermore, the Janus nature of the CrSBr monolayer allows the formation of two inequivalent interfaces for each stacking configuration, providing an intrinsic route to tune the electronic and optical properties of the heterostructure through interface selection.

Motivated by the unique and complementary properties of non-Janus MoTe$_2$ and Janus CrSBr, and the additional functionality that emerges from their integration, we investigate the electronic, optical, and excitonic properties of the MoTe$_2$/CrSBr vdW heterostructure using density functional theory (DFT) and many-body perturbation theory (MBPT). Quasiparticle energies are computed within the fully relativistic single-shot GW approximation, and excitonic effects are evaluated by solving the Bethe-Salpeter equation (BSE). We construct two interface configurations corresponding to Te-S and Te-Br contacts and show that both are dynamically and thermally stable and exhibit type-II band alignment with a direct band gap, enabling the formation of long-lived interlayer excitons. Owing to the intrinsic polarity of the Janus layer, the two interfaces generate distinct built-in electric fields that strongly modulate the quasiparticle band gap, optical gap, exciton binding energy, and exciton radiative lifetime. These results demonstrate that interface-dependent electrostatic potentials provide an effective route for tuning excitonic properties, highlighting the MoTe$_2$/CrSBr heterostructure as a promising candidate for optoelectronic applications.

## II. COMPUTATIONAL METHODS

The DFT calculations were performed using the QUANTUM ESPRESSO package [51, 52]. The exchange-correlation interactions were described within the Perdew-Burke-Ernzerhof (PBE) functional [53], where the valence electrons were modeled using the optimized norm-conserving Vanderbilt pseudopotentials from the PseudoDojo library [54]. To accurately capture SOC effects, fully relativistic pseudopotentials were used. The plane-wave energy cutoff was set to 80 Ry and the vdW interactions were included using the DFT-D3 correction scheme [55]. A vacuum space of 40 Å was introduced to eliminate interactions between periodic images along the non-periodic direction and a truncated Coulomb potential (TCP) [56] with a box-like cutoff was applied to prevent artificial screening effects. A 24×24×1 Monkhorst-Pack k-point grid was used in the Brillouin Zone (BZ) sampling, ensuring total-energy convergence within $10^{-3}$ Ry.

To obtain quantitative predictions of the electronic structure, the quasiparticle energies were calculated within the single-shot GW approximation [57, 58], and the the screened Coulomb interaction was treated using the plasmon-pole approximation (PPA) for all k-points in the BZ [59]. The energy cutoffs were set to 80 Ry for the exchange part of the self-energy and 30 Ry for the response function. Convergence of the GW band gap was achieved using 200 virtual bands in the polarization function and self-energy and the excitonic properties were calculated by solving the BSE on top of the GW quasiparticle energies, including 10 valence and 10 conduction bands.

The GW and BSE results were calculated using the YAMBO code [60]. To greatly accelerate the convergence of the GW band gap with respect to the number of virtual bands in the polarization function and self-energy, the extrapolar correction scheme proposed by Bruneval and Gonze (BG) [61] was implemented. The random integration method (RIM) was also used to remove numerical divergences in the evaluation of Coulomb integrals through Monte Carlo sampling. In these calculations, $10^6$ random q-points and 200 random G-vectors were included to ensure numerical accuracy. Moreover, the stochastic integration of the screened potential (SISP) scheme [62] was implemented to accelerate the convergence of the GW and BSE calculations with respect to the number of k-points in the BZ.

The exciton radiative lifetime was calculated using the Fermi's Golden rule [63, 64]. At $T = 0$ K, the lifetime was obtained as:

$$\tau_S(0) = \frac{\hbar^2 cA}{8\pi e^2 E_s(0)\mu_S^2},$$

where $\hbar$ is the reduced Planck constant, $c$ is the speed of light, $A$ is the area of unit cell, $E_S(0)$ is the exciton excitation energy, and $\mu_S$ is the transition dipole moment of the exciton. The transition dipole moment is the electric dipole moment associated with a transition between an initial state $a$ and a final state $b$, and is defined as:

$$\mu_S = \langle\psi_b|r|\psi_a\rangle = \frac{i\hbar}{(E_b - E_a)m}\langle\psi_b|P|\psi_a\rangle,$$

where $\psi_a$ and $\psi_b$ are eigenstates with energy $E_a$ and $E_b$, respectively, and $m$ denotes the electron mass.

## III. RESULTS AND DISCUSSION

### A. MoTe$_2$ and CrSBr isolated monolayers

Figure 1 shows the top and side views of the isolated MoTe$_2$ and CrSBr monolayers. Both monolayers adopt a hexagonal lattice; MoTe$_2$ crystallizes in the 2H phase, whereas CrSBr exhibits a 1T phase. The optimized lattice constants are 3.55 Å for MoTe$_2$ and 3.52 Å for CrSBr, resulting in a small lattice mismatch of less than 1%. This close structural compatibility minimizes strain effects upon stacking and supports the formation of a stable vdW heterostructure. Despite their similar in-plane lattice symmetry, the two monolayers exhibit distinct out-of-plane symmetry characteristics. The MoTe$_2$ monolayer preserves out-of-plane mirror symmetry, owing to the chemically identical Te layers symmetrically arranged around the central Mo plane, whereas CrSBr monolayer lacks mirror symmetry due to its Janus nature, with S and Br atoms occupying opposite sides of the Cr layer. This intrinsic structural asymmetry in CrSBr gives rise to a built-in out-of-plane dipole, which is expected to play a critical role in determining the interfacial electrostatic potentials and interface-dependent properties of the MoTe$_2$/CrSBr heterostructure. While not discussed here, this out-of-plane dipole will also enable 2D piezoelectric behavior and response, highlighting the versatility of layered CrSBr [65, 66]. The optimized lattice constant of the MoTe$_2$ monolayer is in excellent agreement with previous theoretical [67] and experimental reports [68].

The MoTe$_2$ monolayer has been experimentally synthesized and is known to exhibit high ambient stability [43, 69, 70]. In contrast, the structural stability of the proposed hexagonal CrSBr monolayer is systematically examined in this work. The stability was assessed through both lattice-dynamical and finite-temperature analyses, as summarized in Figure 2. The phonon dispersion spectrum shown in Figure 2(a) exhibits no imaginary frequencies throughout the BZ, indicating the absence of soft modes and confirming the dynamical stability of the monolayer. To further evaluate its thermal stability, *ab initio* molecular dynamics (AIMD) simulations were performed at 300 K for 5 ps. As shown in Figure 2(b), the total energy exhibits only small fluctuations around its equilibrium value, and no noticeable structural distortions are observed in the final atomic configuration. These results demonstrate that the hexagonal CrSBr monolayer is both dynamically and thermally stable, supporting its suitability as a constituent layer in the proposed vdW heterostructure.

Figure 3(a-d) presents the atom-projected electronic band structures of the isolated MoTe$_2$ and CrSBr monolayers, resolved into spin-down and spin-up components. For MoTe$_2$, the spin-down and spin-up band structures are identical, confirming its non-magnetic nature [47]. The material exhibits a direct band gap of 1.08 eV, with both the valence band maximum (VBM) and conduction band minimum (CBM) located at the K point. The calculated gap is in agreement with the previously reported values of 1.11 eV [71] and 1.00 eV [72]. The atom-projected bands further indicate that the band edges are predominantly derived from Mo atoms, with smaller contributions from Te atoms, particularly in the conduction band. In contrast, CrSBr exhibits a spin-dependent electronic structure characteristic of a magnetic semiconductor [73]. In the spin-down channel, the VBM is located at the Γ point, while the CBM appears at the M point, resulting in an indirect band gap of 2.32 eV. For the spin-up channel, the VBM lies along the Γ-K path, whereas the CBM is located at the K point, yielding a smaller indirect band gap of 1.03 eV. Consequently, the fundamental band gap of the CrSBr monolayer is governed by the spin-up component. The atom-projected band structure shows that, in the spin-up channel, the electronic states near both band edges of the CrSBr monolayer are dominated by Cr-derived states, with only minor contributions from S and Br atoms. These results highlight the spin-degenerate nature of the MoTe$_2$ monolayer, in contrast to the spin-resolved electronic structure of the CrSBr monolayer, where the band edges and band gaps differ noticeably between the two spin channels.

Figure 4 depicts the relative band-edge positions of the isolated $MoTe_2$ and CrSBr monolayers, referenced to the vacuum level. The CBM and VBM of $MoTe_2$ are both located at higher (less negative) energies than the corresponding band edges of CrSBr, establishing a staggered type-II band alignment [35]. Specifically, the CBM of $MoTe_2$ is positioned at -2.70 eV, whereas the CBM of CrSBr lies at -3.15 eV, yielding a noticeable conduction-band offset that energetically favors electron transfer from the $MoTe_2$ to the CrSBr layer. Similarly, the VBM of $MoTe_2$ is located at -3.78 eV, while the VBM of CrSBr is found at -4.18 eV, producing a sizable valence-band offset that drives hole localization in the $MoTe_2$ layer. Such band alignment illustrates the spatial separation of electrons and holes across the interface, with electrons residing in CrSBr and holes in $MoTe_2$, indicating favorable conditions for the formation of long-lived interlayer excitons in the $MoTe_2$/CrSBr heterostructure.

Figure 5 presents the optical absorption spectra and quasiparticle electronic band structures of the isolated $MoTe_2$ and CrSBr monolayers calculated at the GW+BSE level including SOC interaction. As the $MoTe_2$ monolayer exhibits strong SOC effects [44], all quasiparticle and optical calculations were carried out at the GW+BSE level with SOC included. For $MoTe_2$, the imaginary part of the dielectric function in Figure 5(a) exhibits the first absorption peak at 1.14 eV, which is associated with a bright exciton due to its large oscillator strength. By correlating this feature with the GW+SOC band structure in Figure 5(b), this excitonic transition is identified as originating from direct vertical transitions between the SOC-split valence- and conduction-band edges at the $K^+$ and $K^-$ valleys. Using the quasiparticle gap of 1.65 eV, the resulting exciton binding energy—evaluated as the difference between the quasiparticle and optical gaps—is 0.51 eV, in close agreement with the previously reported values of 0.46 eV [46] and 0.58 eV [50], reflecting the strong Coulomb interaction and reduced dielectric screening in this monolayer. The calculated GW gap also agrees well with the reported value of 1.72 eV [46, 50], confirming the accuracy of our results. The optical parameters are provided in Table I. The second and third peaks of the optical spectrum appear at 1.44 eV and 1.80 eV, respectively, and involve vertical transitions between deeper valence bands and higher conduction bands near the K valleys. These peaks also correspond to bright excitons, as identified by their significant oscillator strengths. The GW+SOC band structure further reveals a substantial SOC-induced splitting of approximately 220 meV at the valence-band edge and a smaller splitting of 34 meV at the conduction-band edge, consistent with the strong SOC effects associated with the heavy Mo and Te atoms and the lack of inversion symmetry in the monolayer. Our results are consistent with those reported previously [45].

In contrast, the optical response of the CrSBr monolayer, shown in Figure 5(c), is dominated by a strong absorption peak centered at 1.55 eV, while the lowest-energy bright exciton ($X_{1b}$) appears at 1.43 eV and defines the optical gap of the material. Comparison with the GW+SOC band structure in Figure 5(d) indicates that this exciton originates from direct vertical transitions along the Γ-$K^+$ and Γ-$K^-$ paths, where the highest valence band and lowest conduction band approach each other in energy. Using the smallest direct quasiparticle gap of 2.40 eV, the corresponding exciton binding energy is extracted as 0.97 eV, substantially larger than in $MoTe_2$ (see Table I), highlighting the enhanced electron-hole interaction arising from the more localized, Cr-derived electronic states and weaker dielectric screening in CrSBr monolayer. Despite the inclusion of SOC interaction, no appreciable SOC-induced splitting is resolved at the band edges, reflecting the weaker relativistic effects associated with the lighter constituent elements. From the GW+SOC band structure one can also identify an indirect band gap of 2.01 eV. All the optical parameters are listed in Table I. The different electronic and excitonic characteristics of the $MoTe_2$ and CrSBr monolayers highlight the potential of their integration into a vdW heterostructure for enabling rich and tunable optoelectronic properties. $MoTe_2$ supports SOC-split bright excitons at the K valleys, whereas CrSBr hosts strongly bound bright excitons arising from direct transitions along the Γ-K direction, reflecting their contrasting spin-orbit and Coulomb interaction regimes. This complementarity provides a favorable platform for engineering interlayer charge separation, long-lived excitonic states, and interface-dependent optical responses, making the $MoTe_2$/CrSBr heterostructure a promising candidate for advanced optoelectronic and photovoltaic applications.

### B. $MoTe_2$/CrSBr heterostructure

After establishing the electronic and optical properties of the isolated monolayers, we assess the structural stability of the $MoTe_2$/CrSBr heterostructure. To identify the energetically favorable stacking configuration, a set of 36 high- and low-symmetry configurations were systematically explored by mapping the potential energy surface. As shown in Figure 6(a), the minimum potential energy corresponds to the high-symmetry AA stacking, indicating that this configuration is the most stable. In this geometry, the Cr atoms are aligned above the Mo sites, the S atoms are positioned above the Te atoms, and the Br atoms occupy hollow sites of the $MoTe_2$ hexagonal lattice. The inset of Figure 6(a) shows the corresponding side view of the AA stacking. Consequently, all subsequent analyses are restricted to this stacking configuration.

Due to the intrinsic out-of-plane asymmetry of the Janus CrSBr monolayer, two different interfaces naturally emerge within the AA stacking configuration, namely the Te-S and Te-Br interfaces, depending on whether the S or Br sublayer faces the $MoTe_2$ sheet. In the Te-S interface, the sulfur side of CrSBr faces the $MoTe_2$ layer, while in the Te-Br interface, the bromine side of CrSBr interacts with $MoTe_2$. The dynamical stability of both interfaces was evaluated through their phonon dispersion spectra, shown in Figures 6(b) and 6(c). The absence of imaginary phonon modes throughout the BZ confirms that both heterostructures are dynamically stable.

To further examine their thermal stability, the AIMD simulations were performed at 300 K for 5 ps, and the corresponding total energy profiles are presented in Figures 6(d) and 6(e). In both cases, the total energy exhibits only moderate fluctuations around the equilibrium value, and no structural degradation or bond breaking is observed in the final atomic configurations. These results demonstrate that both the Te-S and Te-Br interfaces of the AA stacking are thermally stable under ambient conditions, supporting the feasibility of experimentally realizing the $MoTe_2$/CrSBr heterostructure and providing a reliable structural foundation for the interface-dependent electronic and excitonic phenomena discussed below.

The planar-averaged electrostatic potential profiles of the $MoTe_2$/CrSBr heterostructure are presented in Figures 7(a) and 7(c). A significant potential drop is observed across the interface, indicating the presence of a strong built-in electric field that cannot be neglected. For the Te-Br interface, the potential difference extracted between the Te and Br atomic planes reaches 9.017 eV, which is significantly larger than the corresponding value of 6.256 eV for the Te-S interface, implying a stronger built-in electric field in the Te-Br configuration. In both cases, the electric field points from the CrSBr layer toward the $MoTe_2$ layer, due to the deeper electrostatic potential of the $MoTe_2$. Using the interlayer distances listed in Table II, the resulting built-in electric fields are estimated to be 1.64 eVÅ$^{-1}$ for the Te-S interface and 2.66 eVÅ$^{-1}$ for the Te-Br interface. The stronger built-in electric field across the Te-Br interface enhances interlayer charge separation by driving electrons and holes into spatially separated layers, thereby suppressing electron-hole recombination and stabilizing long-lived interlayer excitons. The calculated built-in electric fields in the $MoTe_2$/CrSBr heterostructure are significantly stronger than those reported for conventional

TMD heterostructures [74-76], enabling more effective charge separation, enhanced exciton dissociation rate, and improved performance of optoelectronic devices.

Bader charge analysis further reveals a charge transfer of approximately $1.82\times10^{-2}$ e for the Te-S interface and $2.40\times10^{-2}$ e for the Te-Br interface from the MoTe$_2$ layer to the CrSBr layer, demonstrating noticeable charge polarization. The corresponding charge density difference plots shown in Figures 7(b) and 7(d) exhibit clear electron accumulation in the CrSBr layer and electron depletion in the MoTe$_2$ layer, supporting the Bader charge analysis. This interfacial charge redistribution generates a strong polarization field localized at the interface, which promotes efficient separation of electrons and holes into the CrSBr and MoTe$_2$ layers and is beneficial for optoelectronic and photovoltaic applications [16]. Based on the calculated vacuum energy level, the work function of the heterostructure is determined to be 4.50 eV for the Te-S interface and 5.02 eV for the Te-Br interface.

Figure 8 demonstrates the layer-projected electronic band structures of the MoTe$_2$/CrSBr heterostructure for both spin-down and spin-up channels. For the Te-S interface, in the spin-down channel, the heterostructure is a semiconductor with the VBM located at the K point and the CBM at the M point, resulting in an indirect band gap of 0.39 eV. On the contrary, the spin-up channel exhibits a direct band gap of 0.14 eV, with both the VBM and CBM located at the K point, indicating that the fundamental band gap of the Te-S interface is governed by the spin-up component. The layer-projected bands clearly show that the VBM is derived from the MoTe$_2$ layer, while the CBM originates mainly from the CrSBr layer, confirming the type-II band alignment identified in Figure 4. For the Te-Br interface, in the spin-down channel, the heterostructure remains an indirect-gap semiconductor, with the VBM located at K and the CBM at M, but with a significantly larger band gap of 0.87 eV. Similarly, in the spin-up channel, a direct band gap of 0.62 eV is obtained at the K point, which again defines the fundamental gap of the heterostructure. Similar to the Te-S case, the layer-resolved projections indicate that the VBM is dominated by MoTe$_2$ states, whereas the CBM is primarily contributed by CrSBr states, indicating that the type-II band alignment is preserved for both interfaces. Overall, both Te-S and Te-Br interfaces of the MoTe$_2$/CrSBr heterostructure display clear spin-dependent electronic behavior, reflecting the spin-polarized nature of the CrSBr layer. As a result, the electronic structure is not spin-degenerate: the positions of the valence- and conduction-band edges, as well as the fundamental band gaps, differ between the spin-up and spin-down channels. All the electronic parameters are summarized in Table II.

The noticeable difference between the band gaps of the Te-S (0.14 eV) and Te-Br (0.62 eV) interfaces originates from the intrinsic out-of-plane polarity of the Janus CrSBr monolayer, which gives rise to interface-dependent dipoles and electrostatic potentials. As shown in Figure 7, CrSBr monolayer itself hosts an internal electric field directed from the Br-terminated side toward the S-terminated side. In the Te-S configuration, this intrinsic electric field aligns with the interfacial electric field of the heterostructure, whereas in the Te-Br configuration the two fields oppose each other. Owing to the higher electronegativity of Br relative to S, the Te-Br interface generates a stronger net interfacial dipole and, consequently, a larger built-in electric field across the junction. This stronger built-in electric field reduces the interlayer valence- and conduction-band offsets, resulting in a larger band gap for the Te-Br interface. In contrast, the weaker dipole at the Te-S interface produces a smaller electrostatic potential difference, leading to a larger interlayer band offset and a correspondingly smaller band gap. Overall, the heterostructure band gap scales directly with the electrostatic potential difference across the interface. The type-II band alignment observed in our results is further supported by the real-space charge-density distributions shown in Figure 8, where the spin-up VBM is localized within the $MoTe_2$ layer, while the corresponding spin-up CBM resides in the CrSBr layer for both interfaces.

The imaginary part of the dielectric function and the corresponding GW+SOC quasiparticle band structures of the $MoTe_2$/CrSBr heterostructure for the Te-S and Te-Br interfaces are presented in Figure 9. For the Te-S interface, the optical spectrum in Figure 9(a) exhibits the lowest-energy bright exciton ($X_{1b}$) at 0.47 eV, which defines the optical gap of the heterostructure and is substantially red-shifted relative to the first bright excitons of the isolated $MoTe_2$ (1.14 eV) and CrSBr (1.43 eV) monolayers. This excitonic resonance originates from direct vertical transitions between the $MoTe_2$-derived VBM and the CrSBr-derived CBM at the $K^+$ and $K^-$ valleys, as indicated by the red arrows in Figure 9(b), and therefore corresponds to an interlayer exciton with the hole localized in the $MoTe_2$ layer and the electron residing in the CrSBr layer. Using the GW+SOC direct band gap of 0.75 eV, the resulting exciton binding energy is estimated to be 0.28 eV, noticeably smaller than those of the constituent monolayers (0.51 eV for $MoTe_2$ and 0.97 eV for CrSBr). This substantial reduction reflects the increased dielectric screening and spatial separation of charge carriers across the interface, which are favorable for suppressing radiative recombination and promoting efficient charge extraction in optoelectronic and photovoltaic devices.

At higher photon energies, additional bright excitonic peaks emerge at approximately 0.86, 1.20, 1.62, and 2.01 eV, followed by a broad and intense absorption feature near 3.0 eV. In particular, the peak at 0.86 eV ($X_{2b}$) is associated with transitions from the MoTe$_2$-derived valence band (VBM-1) to the next CrSBr-derived conduction band (CBM+1) at the $K^+/K^-$ valleys, confirming its interlayer character. Together, these features indicate that the MoTe$_2$/CrSBr heterostructure hosts numerous optically active excitons from the near-infrared to the visible range, extending the absorption window beyond that of the isolated monolayers. The corresponding quasiparticle band structure in Figure 9(b) shows a direct band gap of 0.75 eV at the $K^+/K^-$ points, which is strongly renormalized relative to the monolayers (1.65 eV for MoTe$_2$ and 2.40 eV for CrSBr). A noticeable SOC-induced splitting of about 214 meV appears in the VBM, inherited from the strong SOC of the MoTe$_2$ layer, while the CBM remains essentially unsplit due to its dominant CrSBr character.

For the Te-Br interface, the overall excitonic structure remains similar, but the peaks energies differ quantitatively from those of the Te-S interface. As shown in Figure 9(c), the first bright exciton ($X_{1b}$) appears at 0.84 eV and directly defines the optical gap of the heterostructure. This exciton originates from direct transitions between the MoTe$_2$-derived VBM and the CrSBr-derived CBM at $K^+/K^-$ points, as illustrated in Figure 9(d), confirming its interlayer nature. Using the GW+SOC direct gap of 1.35 eV, the corresponding exciton binding energy is estimated to be 0.51 eV, significantly larger than that of the Te-S interface (see Table III), highlighting the strong sensitivity of exciton binding energy to the interfacial dipole and electrostatic potential imposed by the Janus CrSBr monolayer. Additional bright features emerge at around 1.40 eV and 2.47 eV, with the former primarily associated with intralayer transitions within the MoTe$_2$ component, indicating a crossover from interlayer to intralayer excitonic character at higher energies. The quasiparticle band structure of the Te-Br interface shows a direct gap at $K^+/K^-$, accompanied by a SOC-induced splitting of about 215 meV at the VBM, while the CBM remains essentially spin-degenerate due to its CrSBr-derived character.

The presence of type-II band alignment in both Te-S and Te-Br interfaces indicates that the lowest-energy excitations in the MoTe$_2$/CrSBr heterostructure are spatially separated interlayer excitons, which are expected to exhibit longer radiative lifetimes than the intralayer excitons of the isolated monolayers [34]. Such extended lifetimes are particularly advantageous for optoelectronic applications [36], as they provide a longer time window for exciton dissociation into free charge carriers prior to recombination. To quantify this effect, we calculated the radiative lifetimes of the first bright excitons for the MoTe$_2$ and CrSBr monolayers as well as for the MoTe$_2$/CrSBr heterostructure with both Te-S and Te-Br interfaces. The results are summarized in Tables I and III. For the isolated monolayers, the exciton lifetime is found to be 3.6 ps for MoTe$_2$, in close agreement with previously measured value (3.4 ± 0.5 ps) [46], while CrSBr exhibits a longer lifetime of 8.1 ps, reflecting difference in transition dipole moment.

In contrast, the heterostructure shows a noticeable enhancement of the radiative lifetime due to the reduced electron-hole overlap associated with interlayer excitons. Specifically, the exciton lifetime increases to 18.7 ps for the Te-S interface and 44.8 ps for the Te-Br interface, both substantially exceeding those of the constituent monolayers. The longer lifetime in the Te-Br configuration can be attributed to a smaller transition dipole moment and a stronger built-in electric field, which further separates the electron and hole across the two layers and suppresses radiative recombination. Particularly, these values are also larger than those reported for several well-known TMD heterostructures, such as MoS$_2$/WS$_2$ (12.3 ps) [64], MoSe$_2$/WSe$_2$ (8.7 ps) [64], and MoSe$_2$/WSSe (18.04 ps) [28]. This significant lifetime enhancement confirms that the MoTe$_2$/CrSBr heterostructure provides an efficient platform for achieving long-lived interlayer excitons, reinforcing its suitability for applications that rely on effective charge separation and reduced recombination losses.

Further, the unique combination of a non-centrosymmetric, magnetic, and optically active heterostructure opens up a range of potential applications which can be studied in the future. The non-centrosymmetric nature enables piezoelectric and pyroelectric (mechano- and thermo-responsive) properties coupled with magnetic or optical readout. This type of responsive system can be considered analogous to nitrogen vacancies in diamond for quantum sensing [77] or pseudo-multiferroic applications for edge computing [78]. Further, these properties occurring in a commensurate stacking structure (AA) provide an advantage over metastable twistronic architectures exploiting such emergent phenomena in 2D vdW heterostructures [79, 80]. To achieve these properties, this heterostructure must be experimentally realized, which is currently limited by challenges in the stabilization of MoTe$_2$ due to the volatility of Te [81], and the lack of reproducible approaches to produce monolayer CrSBr, with current methods producing bulk crystals via chemical vapor transport and subsequent mechanical exfoliation [82, 83]. Nevertheless, this work highlights the immense potential of heterostructure formation between TMDs and CrSBr.

## Conclusion

In conclusion, we have carried out a comprehensive investigation of the structural, electronic, optical, and excitonic properties of the MoTe$_2$/CrSBr van der Waals heterostructure by combining density functional theory with fully relativistic GW and Bethe-Salpeter equation calculations. To clearly reveal the role of interlayer coupling, the optoelectronic properties of the isolated MoTe$_2$ and CrSBr monolayers were also examined as reference systems. A systematic exploration of 36 high- and low-symmetry stacking configurations identified AA stacking as the most energetically favorable structure. Owing to the intrinsic out-of-plane asymmetry of the Janus CrSBr monolayer, this stacking gives rise to two different interfaces, namely Te-S and Te-Br. Our results demonstrate that both interfaces are dynamically and thermally stable and exhibit type-II band alignment with a direct quasiparticle gap, enabling efficient spatial separation of electrons and holes across the two layers. This band alignment leads to the formation of interlayer excitons with reduced electron-hole overlap and consequently extended radiative lifetimes. The calculated lifetimes reach approximately 18.70 ps and 44.80 ps for the Te-S and Te-Br interfaces, respectively, significantly exceeding those of the intralayer excitons in the constituent monolayers. At the same time, the two interfaces show different optical characteristics: the optical gap is found to be 0.47 eV for the Te-S interface and 0.84 eV for the Te-Br interface, while the corresponding exciton binding energies are 0.28 eV and 0.51 eV. These differences originate from the distinct built-in electric fields generated by the Janus layer, which modify the interfacial electrostatic potential and, in turn, strongly influence quasiparticle energies, exciton binding energy, and recombination dynamics. Overall, our findings reveal that the intrinsic polarity of the Janus CrSBr layer provides an effective and intrinsic route to control excitonic properties through interface selection alone, without the need for external fields or strain engineering. The coexistence of type-II band alignment, spatially separated interlayer excitons, and interface-tunable optical response establishes the MoTe$_2$/CrSBr heterostructure as a promising platform for designing systems with long-lived excitons. These characteristics make it a compelling candidate for future applications in nanoscale optoelectronics, photodetection, and energy-harvesting technologies, where controlled charge separation and extended exciton lifetimes are essential.

## CONFLICT OF INTEREST

The authors have no conflicts to disclose.

## DATA AVAILABILITY

The code for calculation of the DFT results can be found at [https://www.quantum-espresso.org/] with [DOI 10.1088/0953-8984/21/39/395502]. The version of the code employed for this study is version [6.0].

## ACKNOWLEDGMENTS

M. A. M acknowledges the funding support by Iran National Science Foundation (INSF) under project No.4023062. C.A. was supported by the "MagTop" project (FENG.02.01-IP.05-0028/23) carried out within the "International Research Agendas" programme of the Foundation for Polish Science, co-financed by the European Union under the European Funds for Smart Economy 2021-2027 (FENG).

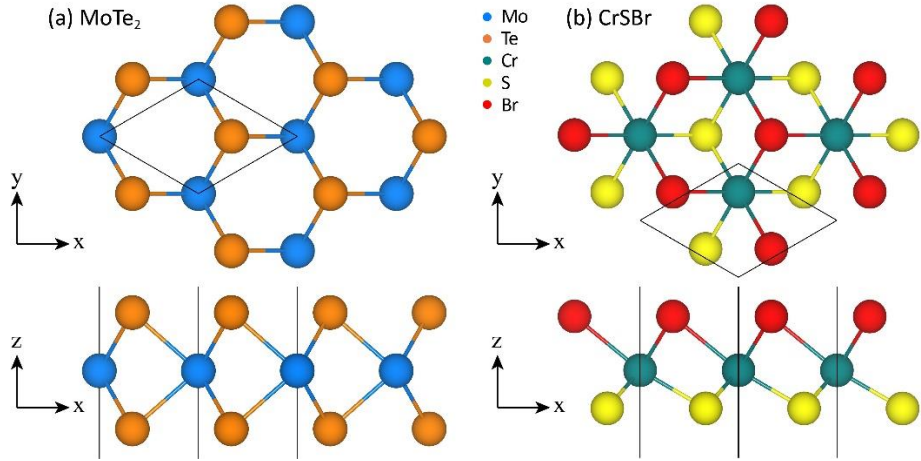

Figure 1. Top and side views of the (a) MoTe$_2$ and (b) CrSBr monolayers. The Mo, Te, Cr, S, and Br atoms are shown in blue, orange, green, yellow, and red, respectively. The solid lines indicate the unit cells.

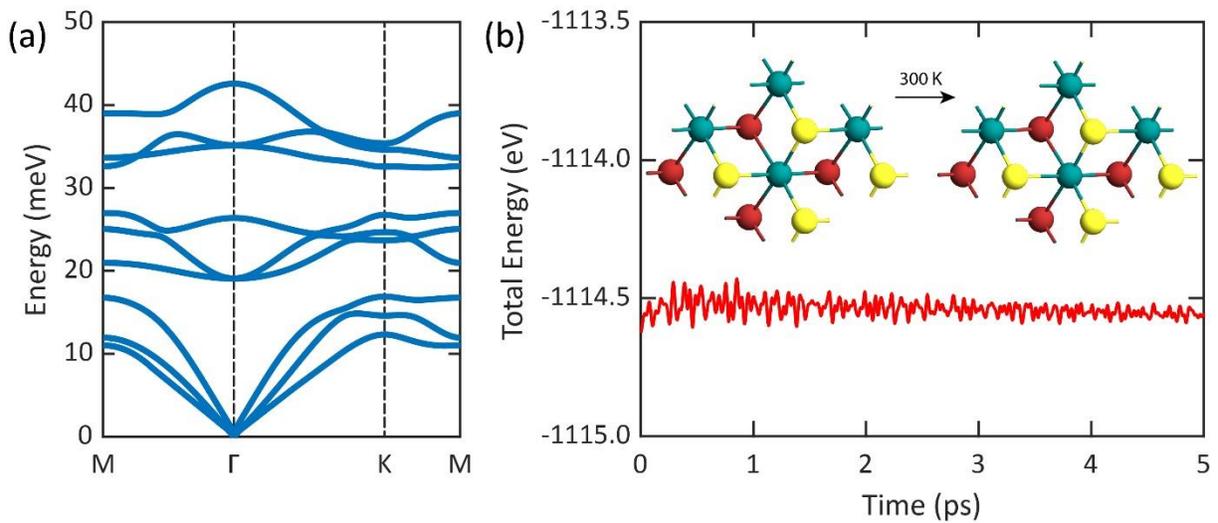

Figure 2. (a) Phonon dispersion spectrum of the CrSBr monolayer. (b) Total energy profile of the CrSBr monolayer during the AIMD simulations at 300 K. The inset shows the atomic structures before and after the simulation. The Cr, S, and Br atoms are shown in green, yellow, and red, respectively.

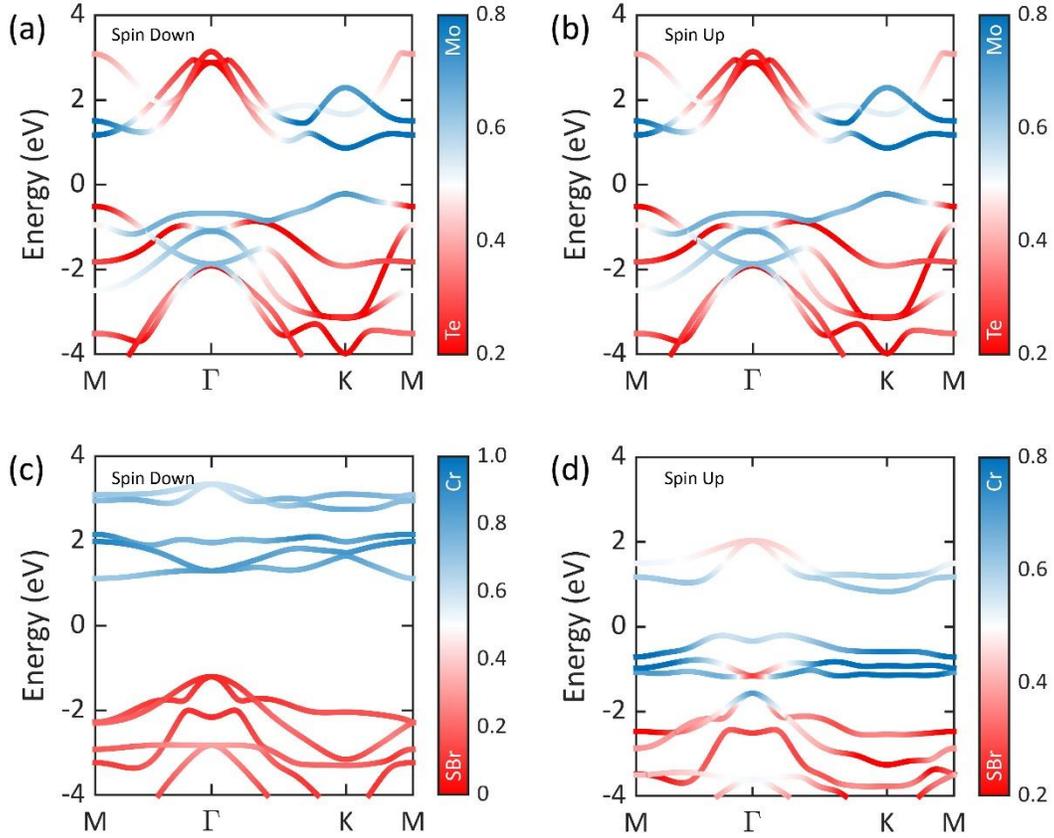

Figure 3. Spin-polarized projected band structure of the MoTe$_2$ (top panel) CrSBr (bottom panel) monolayers for the spin-down (a) and (c) and spin-up (b) and (d) components.

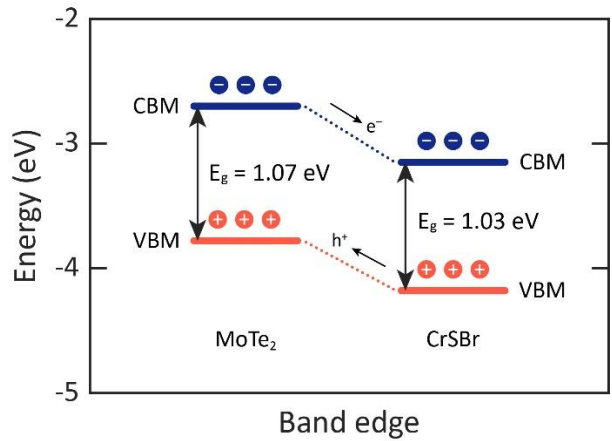

Figure 4. Band edge positions of the MoTe$_2$ and CrSBr monolayers relative to the vacuum level, showing a Type-II band alignment. Such an alignment in direct-gap heterostructures results in the formation of interlayer excitons.

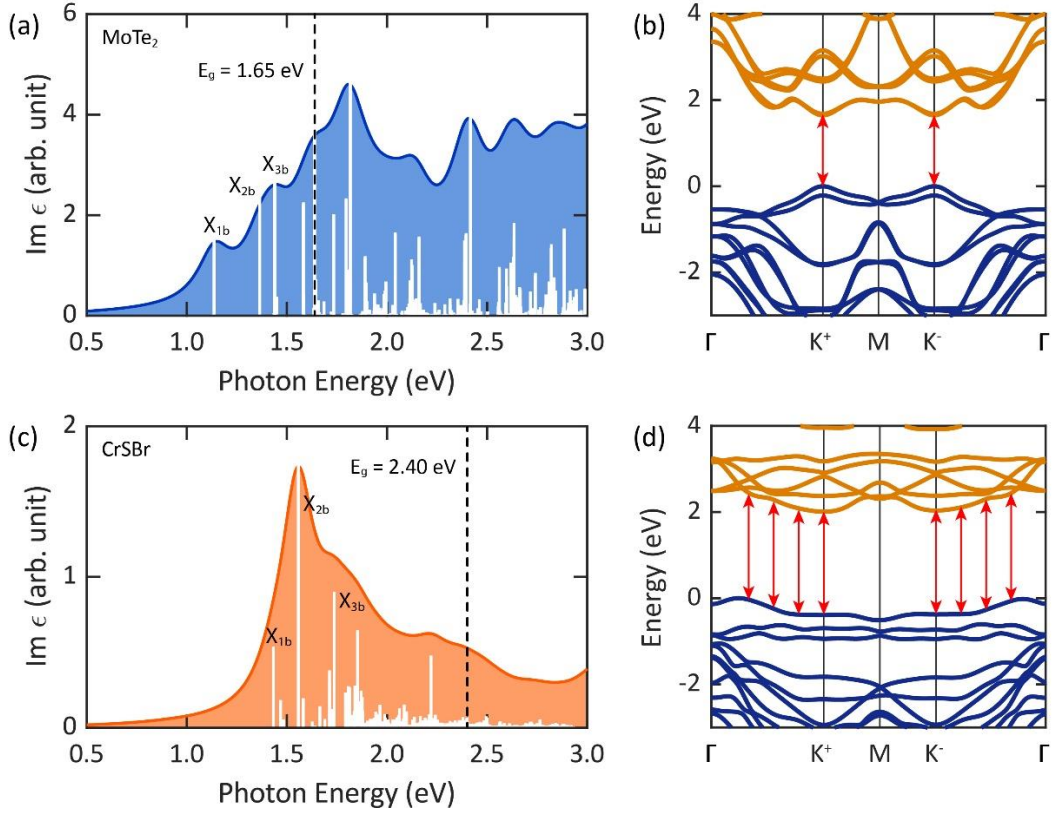

Figure 5. Imaginary part of the dielectric functions (left panel) of the MoTe$_2$ (a) and CrSBr (c) monolayers, calculated at the GW+BSE level including SOC interaction. The white bars indicate the oscillator strength of optical transitions. The first, second, and third bright excitons of the monolayers are shown by X$_{1b}$, X$_{2b}$, and X$_{3b}$, respectively. The quasiparticle band structures (right panel) of the MoTe$_2$ (b) and CrSBr (d) monolayers are shown at the GW+SOC level. The interband transitions associated with the first bright excitons are highlighted by red arrows.

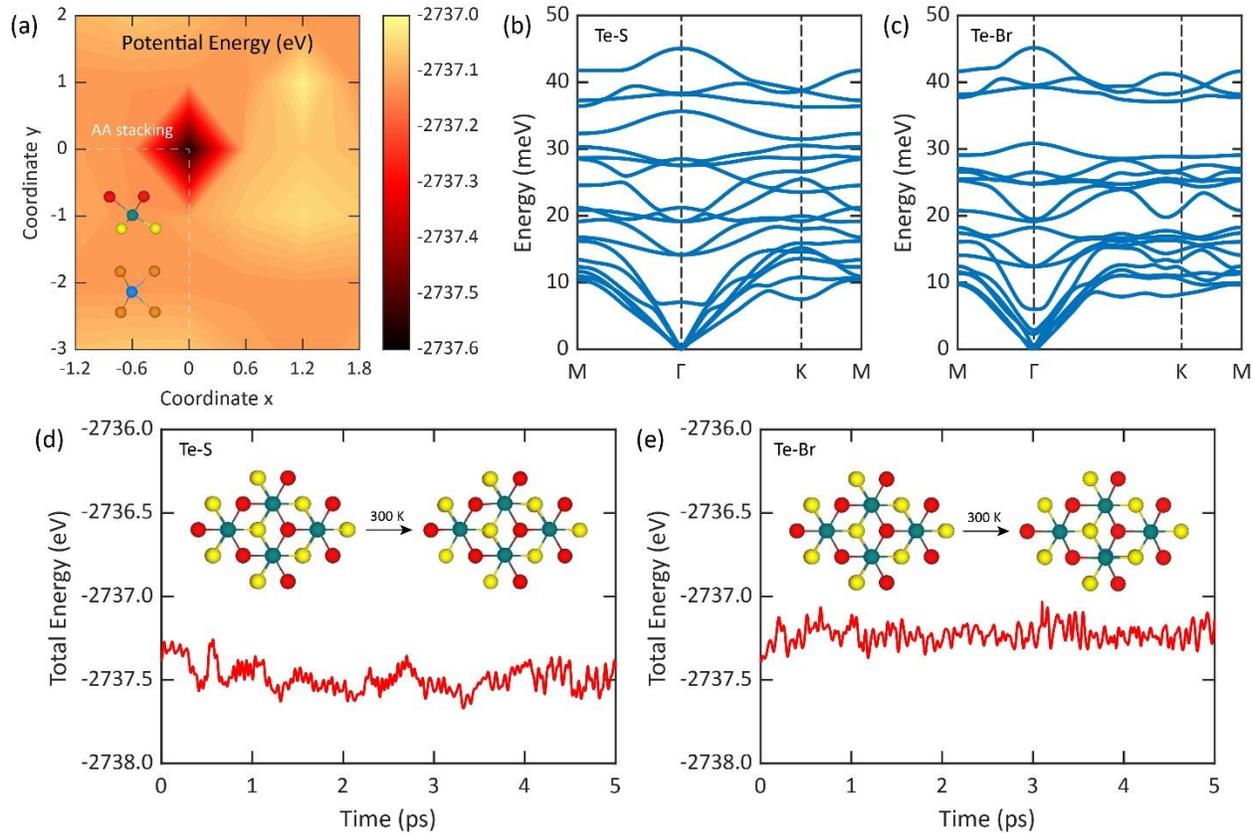

Figure 6. (a) Potential energy surface of the MoTe$_2$/CrSBr vdW heterostructure for all possible stacking configurations, confirming that the AA stacking is the most energetically stable. The inset depicts the side view of this stacking. Phonon dispersion spectrum of the heterostructure for the Te-S (b) and Te-Br (c) interfaces. Total energy profile of the heterostructure for the Te-S (d) and Te-Br (e) interfaces during the AIMD simulations at 300 K. The inset shows the atomic structures before and after the simulation.

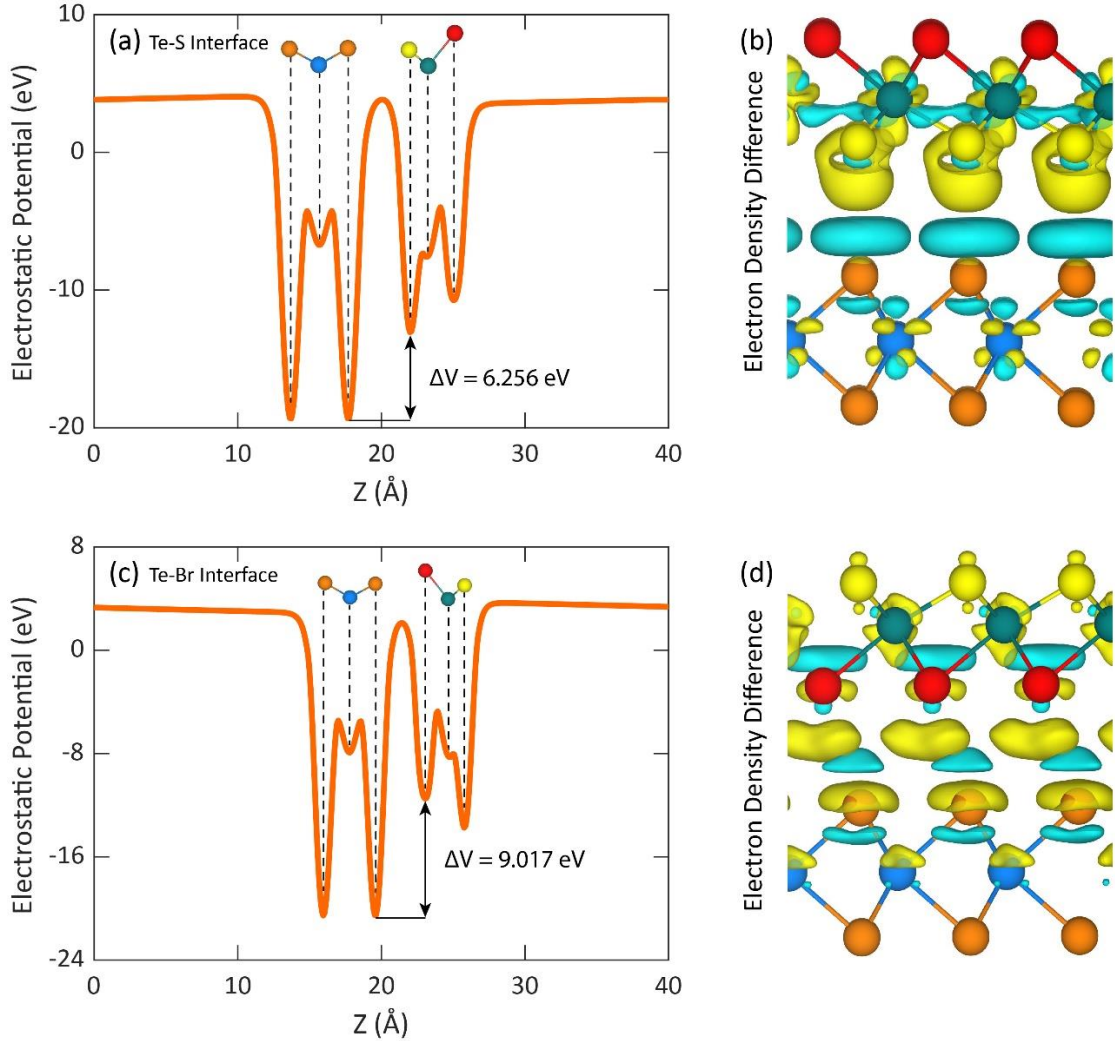

Figure 7. Planar-averaged electrostatic potential profile (left panel) and electron density difference (right panel) of the MoTe$_2$/CrSBr vdW heterostructure for the Te-S (a) and (b) interface and the Te-Br (c) and (d) interface. The potential difference across the Te-Br interface is much larger than that across the Te-S interface, resulting in a stronger built-in electric field. The Mo, Te, Cr, S, and Br atoms are shown in blue, orange, green, yellow, and red, respectively.

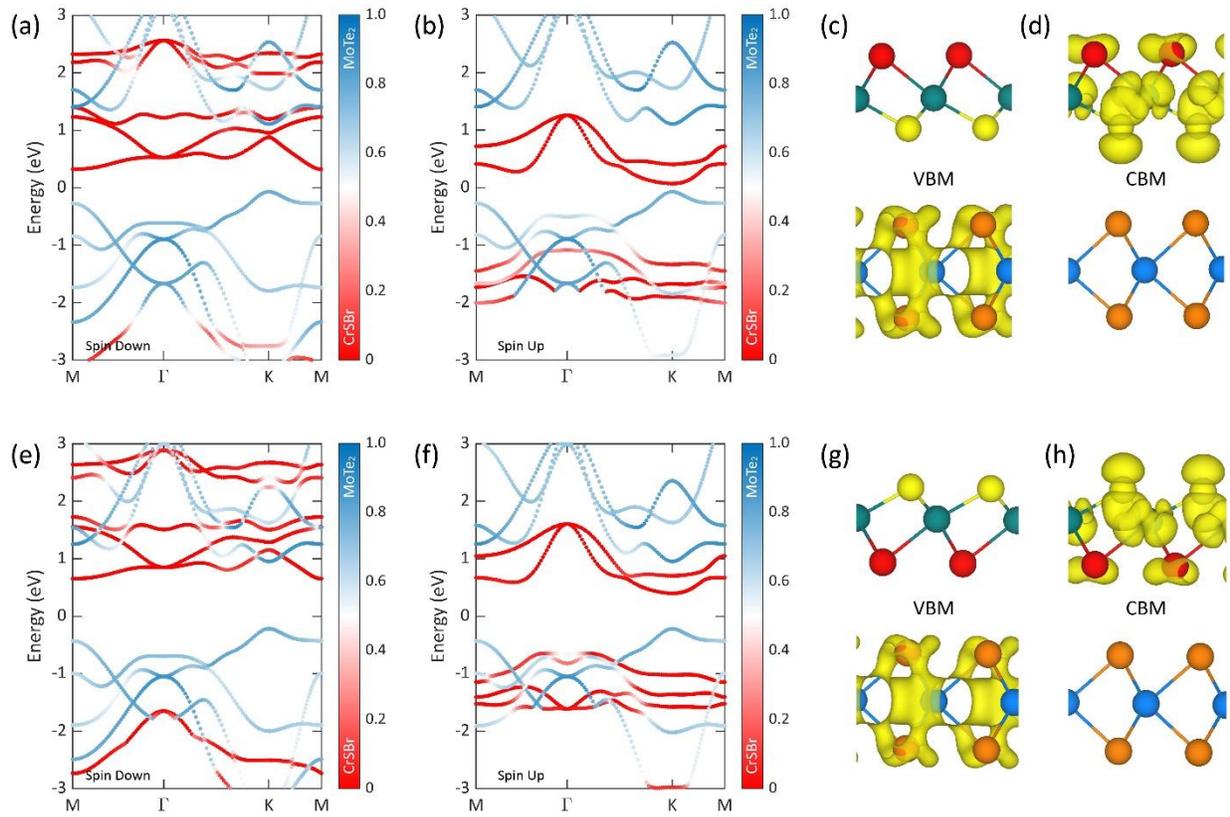

Figure 8. Spin-polarized projected band structure of the MoTe$_2$/CrSBr vdW heterostructure, along with the real-space wave functions of the VBM and CBM, for the Te-S interface (top panel) and the Te-Br interface (bottom panel). The band structures are shown for both the spin-down (a) and (e) and spin-up (b) and (f) components.

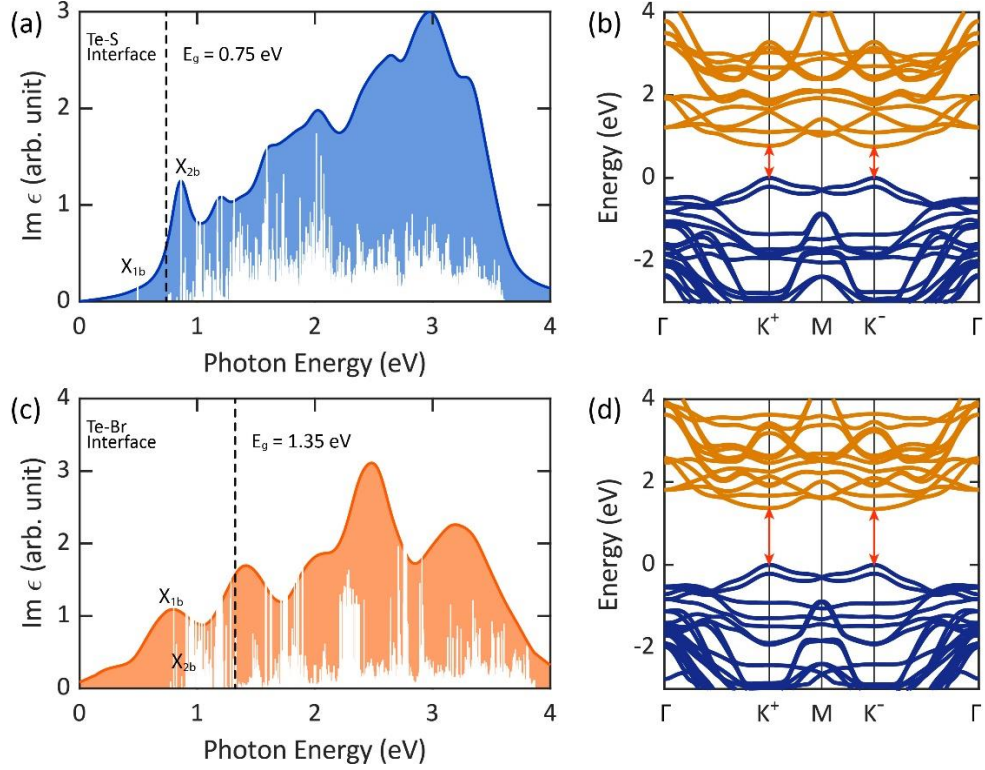

Figure 9. Imaginary part of the dielectric functions (left panel) of the MoTe$_2$/CrSBr vdW heterostructure for the Te-S (a) and Te-Br (c) interfaces, calculated at the GW+BSE level including SOC interaction. The white bars indicate the oscillator strength of optical transitions. The first and second bright excitons of the interfaces are shown by X$_{1b}$ and X$_{2b}$, respectively. The quasiparticle band structures (right panel) of the heterostructure for the Te-S (b) and Te-Br (d) interfaces are shown at the GW+SOC level. The corresponding interband transitions of the first bright excitons are highlighted.

Table I. Structural and electronic parameters of the MoTe$_2$ and CrSBr monolayers, including lattice constants, electronic band gaps (PBE and GW+SOC), locations of the VBM/CBM with SOC, optical gap, exciton binding energy, and exciton radiative lifetime.

| Monolayer | Lattice constant (Å) | PBE/GW band gap (eV) | VBM, CBM | Optical gap (eV) | Exciton binding energy (eV) | Exciton lifetime (ps) |
|---|---|---|---|---|---|---|
| MoTe$_2$ | 3.55 | 1.08 /1.65 | K$^+$, K$^+$ | 1.14 | 0.51 | 3.60 |
| CrSBr | 3.52 | 1.03 /2.01 | G-K$^+$, K$^+$ | 1.43 | 0.97 | 8.09 |

Table II. Structural and electronic parameters of the MoTe$_2$/CrSBr vdW heterostructure for the Te-S and Te-Br interfaces, including lattice constants, interlayer distance, built-in electric field, work function, electronic band gaps (PBE and GW+SOC), locations of the VBM/CBM with SOC, and band alignment.

| Interface | Lattice constant (Å) | Interlayer distance (Å) | Built-in E-field (eV/Å) | Work function (eV) | PBE/GW band gap (eV) | VBM, CBM | Band alignment |
|---|---|---|---|---|---|---|---|
| Te-S | 3.55 | 3.81 | 1.64 | 4.50 | 0.14 /0.75 | K$^+$, K$^-$ | Type-II |
| Te-Br | 3.55 | 3.39 | 2.66 | 5.02 | 0.62 /1.35 | K$^+$, K$^-$ | Type-II |

Table III. Optical and excitonic parameters of the MoTe$_2$/CrSBr vdW heterostructure for the Te-S and Te-Br interfaces, including the optical gap, exciton binding energy, and exciton radiative lifetime.

| Interface | Optical gap (eV) | Exciton binding energy (eV) | Exciton lifetime (ps) |
|---|---|---|---|
| Te-S | 0.47 | 0.28 | 18.70 |
| Te-Br | 0.84 | 0.51 | 44.80 |